\documentclass[prl,showpacs,superscriptaddress,twocolumn,floatfix,amsmath,amssymb]{revtex4-1}%
\usepackage[dvips]{graphicx}
\usepackage{xspace}
\usepackage{color}
\usepackage[dvipsnames]{xcolor}
\usepackage{array}
\usepackage{hyperref}
\usepackage[capitalize]{cleveref}
\usepackage{multirow}
\usepackage{braket}
\usepackage{tabularx}
\usepackage{amsmath}
\usepackage{makecell}
\usepackage{comment}

\newcommand{\itsection}[1]{\textit{#1}---}

\usepackage[normalem]{ulem}

\newcommand{\projector}{\ensuremath{\mathcal P}}

\usepackage{amstext} %
\usepackage{array}   %
\newcolumntype{L}{>{$}l<{$}} %
\newcolumntype{C}{>{$}c<{$}} %

\newcommand{\rucl}{\mbox{$\alpha$-RuCl\textsubscript{3}}\xspace}

\newcommand{\PC}{\ensuremath{\mathcal C}}
\newcommand{\PB}{\ensuremath{\mathcal B}}
\newcommand{\PA}{\ensuremath{\mathcal A}}
\newcommand{\PopX}{\ensuremath{\mathcal O}}
\newcommand{\PX}{\PopX}
\newcommand{\MX}{\ensuremath{X}}
\newcommand{\MY}{\ensuremath{Y}}
\newcommand{\Krylov}{\ensuremath{L}}

\newcommand{\FD}{F\textsubscript{diag}\xspace}
\newcommand{\FV}{F\textsubscript{vert}\xspace}
\newcommand{\FDV}{\FD and \FV}

\newcommand{\chionep}{\ensuremath{\chi^1_{\parallel}}}
\newcommand{\chitwop}{\ensuremath{\chi^2_{\parallel}}}

\newcommand{\Mp}{\ensuremath{\mathcal M}}

\newcommand{\state}[1]{\ensuremath{#1}}

\begin{document}
\newcommand{\frankfurt}{Institut f\"ur Theoretische Physik, Goethe-Universit\"at Frankfurt,
Max-von-Laue-Strasse 1, 60438 Frankfurt am Main, Germany}
\newcommand{\salem}{Department of Physics and Center for Functional Materials, Wake Forest University, Winston-Salem, North Carolina 27109, USA}

\author{David A. S. Kaib}
\email[]{kaib@itp.uni-frankfurt.de}
\affiliation{\frankfurt}
\author{Marius Möller} %
\affiliation{\frankfurt}
\author{Roser Valent{\'\i}}
\affiliation{\frankfurt}
\date{\today}
\title{%
Nonlinear Spectroscopy as a Magnon Breakdown Diagnosis\\ and its Efficient Simulation}
\begin{abstract}
Identifying quantum spin liquids, magnon breakdown, or fractionalized excitations in quantum magnets is an ongoing challenge due to 
the ambiguity of possible origins of excitation continua occurring in linear response probes. 
Recently, it was proposed that techniques measuring  higher-order response, such as two-dimensional coherent spectroscopy (2DCS), could %
{resolve such ambiguities}.
{Numerically simulating} nonlinear response functions can, however, be computationally very demanding. %
We present an efficient Lanczos-based method %
to compute second-order susceptibilities $\chi^{2}(\omega_t,\omega_\tau)$  directly in the frequency domain. %
Applying this to extended Kitaev models describing \rucl, we find qualitatively different nonlinear responses between intermediate magnetic field strengths and the high-field regime. 
To {put these results into context}, we derive the general 2DCS response of partially-polarized magnets within the linear spin-wave approximation, establishing that $\chi^2(\omega_t,\omega_\tau)$ is restricted to a distinct universal form if the excitations are conventional magnons. 
Deviations from this form, %
{as predicted in our (Lanczos-based) simulations for \rucl}, can hence serve %
in 2DCS experiments as direct criteria to determine whether an observed excitation continuum is of conventional two-magnon type or of different nature. %
\end{abstract}

\maketitle

\begin{figure}
    \centering
    \includegraphics[width=.9\linewidth]{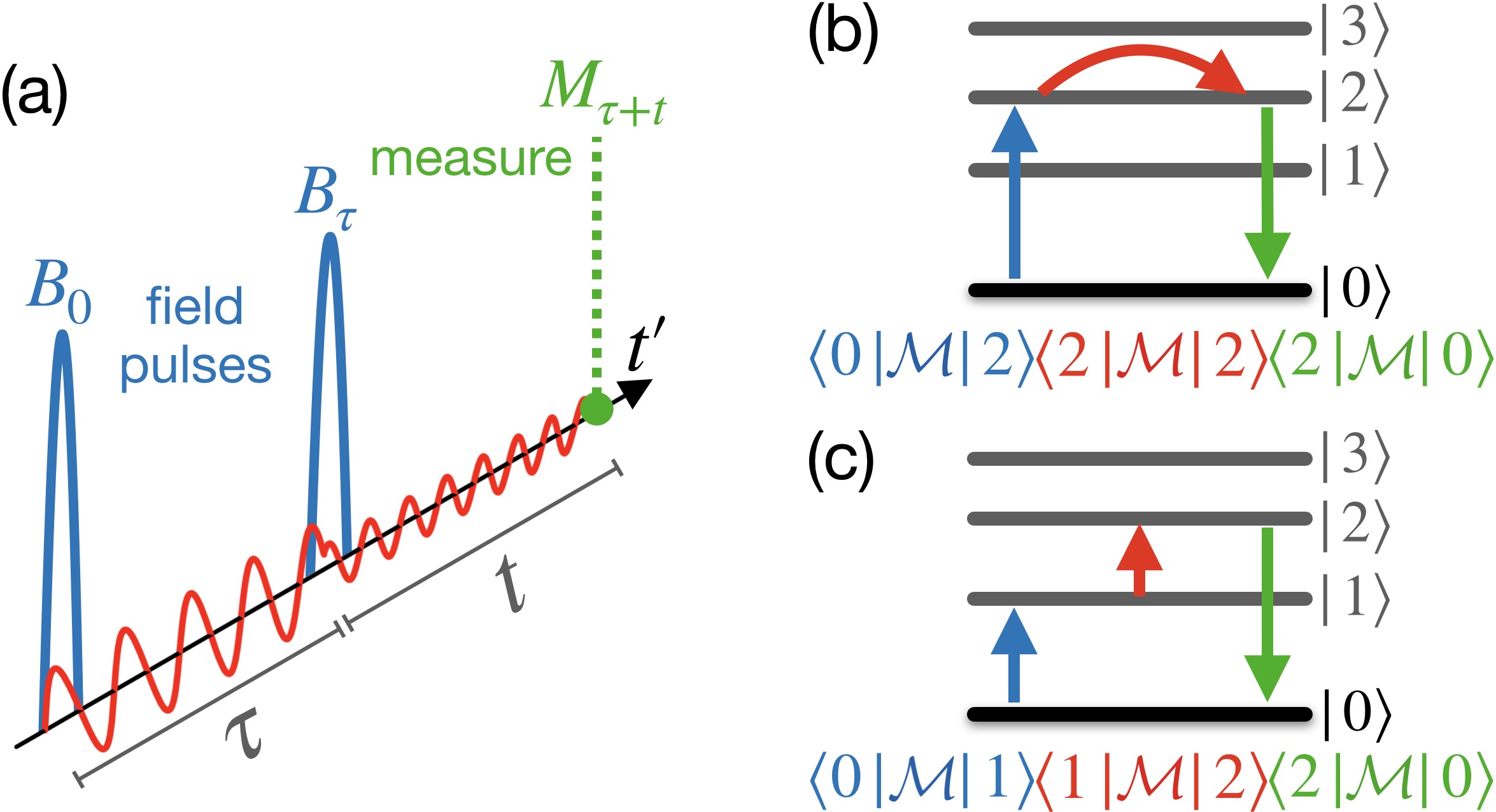}
    \caption{
    (a) Sketch of a 2DCS measurement protocol. Two light pulses are applied with a time difference $\tau$, and the measurement is performed at $t'=\tau+t$ with respect to the first pulse.
    (b,c) Types of matrix elements contributing to zero-temperature second-order susceptibility $\chi^2_{\Mp\Mp\Mp}$.  %
 }
    \label{fig:introduction}
\end{figure}

\itsection{Introduction}%
Nonlinear optics probes such as two-dimensional coherent spectroscopy (2DCS)~\cite{mukamel2000multidimensional} have wide applications in molecular chemistry~\cite{khalil2003coherent,johansson2018nonlinear}, nanomaterials~\cite{sankarnonlinear2018301} and semiconductors \cite{garmirenonlinearsemiconductors1994,kuehn2011TwoDimensionalTerahertzCorrelation}. 
In 2DCS, the time delays between two external field pulses and between measurement are varied [Fig.~\ref{fig:introduction}(a)],
which allows the investigation of higher-order susceptibilities.
Recently, 2DCS has gained much attention in the field of frustrated quantum magnets~\cite{lu2017coherent, choi2020kitaev2dnonlinearspectroscopyforkitaevspinliquidtheory,negahdari2023nonlinear} as a possible highly effective tool for distinguishing quantum spin liquids (QSLs) and other exotic states~\cite{wan2019resolving}. QSLs are generally characterized by absence of magnetic order and the presence of long-range entanglement and fractionalized excitations~\cite{savary2016quantum,knolle2019field}. However, detecting and identifying them experimentally remains a challenge due to a lack
of a smoking gun signature. Moreover, most of their
thermodynamic quantities are quite featureless~\cite{broholm2020quantum}.
Some studies have investigated low-energy fractionalized excitations in QSL candidates by transport measurements, the most prominent being a plateau in thermal Hall measurements in the Kitaev QSL candidate \rucl~\cite{kasahara2018majorana,yokoi2020half}, although such observations are still under debate~\cite{bruin2022robustness,czajka2023planar,lefranccois2023oscillations,dhakal2024spin}. 
In addition to thermal transport, especially the observations of excitation continua  
in {linear response} experiments, have
been taken as evidence for QSL behaviors~\cite{han2012,banerjee2017,wang2017magnetic}. %
Such continua are however difficult to distinguish from continua that can arise, for instance, from two-magnon states or static disorder~\cite{kermarrec2014}. 

2DCS in the terahertz frequency regime, on the contrary, promises to differentiate between {different origins for scattering continua}. 
Reference~\cite{wan2019resolving} demonstrated this by analytically investigating the exactly solvable transverse field Ising chain model (TFIM) for which
higher-order susceptibilities distinctly differentiate between cases of dissipationless spinon excitations, spinon decay and disorder. %
Further analytical studies were performed on different models~\cite{li2021photonecho,ruckriegel2024recursive}, including on the Kitaev honeycomb model~\cite{choi2020kitaev2dnonlinearspectroscopyforkitaevspinliquidtheory,krupnitska2023finitetempSHG,brenig2024response,kanega2021nonlinearopticalresponsesinkitaevspinliquids,qiang2023probing}, 
where the possibility to probe fractionalized excitations in high-harmonic generation probes was shown.
Recently, there have also been first numerical simulations~\cite{sim2023nonlinearboundstatesperturbedisingchains,sim2023nonlinearonedimquantumisingmangets,gao20232dcspectrum,watanabe2024exploring} of higher-order susceptibilities in quantum magnets,
including infinite matrix-product state calculations (IMPS)~\cite{sim2023nonlinearboundstatesperturbedisingchains,sim2023nonlinearonedimquantumisingmangets,gao20232dcspectrum}
and exact diagonalization (ED) studies~\cite{watanabe2024exploring}.
In these numerical investigations, nonlinear responses were simulated first in the time domain by explicit discretized time evolution, while the final results in the frequency domain were obtained by Fourier transforms of the two-dimensional time axes. 

Such simulations %
can in principle follow the actual 2DCS measurement protocol~\cite{wan2019resolving,kuehn2011TwoDimensionalTerahertzCorrelation,Woerner20132dcs}: As displayed in \cref{fig:introduction}(a), two pulses with magnetic field along $\alpha$ at $t'=0$ and  along $\beta$ at $t'=\tau$ are applied, and the magnetization along $\gamma$ is measured at $t'=t+\tau$. Given the amplitudes of the two field pulses, $B_{0}^\alpha$ and $B_{\tau}^\beta$, the time-dependent \textit{nonlinear} magnetization along~$\gamma$, $M^\gamma_\mathrm{NL}=M^\gamma_{B_0,B_\tau}-M^\gamma_{B_0}-M^\gamma_{B_\tau}+M^\gamma$, corresponds to
\begin{equation}
    M^\gamma_\mathrm{NL} (t,\tau) = \chi^2_{\gamma \beta \alpha} (t,\tau+t)\,  B_\tau^\beta B_0^\alpha + \mathcal{O}({B_{t'}}\!^3),
\end{equation} 
giving access to the leading-order nonlinear susceptibility $\chi^2_{\gamma \beta \alpha} (t,\tau+t)$ and %
higher-order susceptibilities contained in $\mathcal O({B_{t'}}\!^3)$. Explicitly simulating this time-dependent experiment is however rather computationally involved. 

In the present work, we introduce an alternative approach to calculate higher-order response functions efficiently in an %
{ED} framework operating directly in the frequency domain. 
We successfully benchmark the method with known results for the transverse-field Ising model.
We then apply this method to extended Kitaev models under magnetic field, relevant to \rucl, and focus on the polarization channel ``$\chitwop$'', whose corresponding linear response ($\chionep$) features an excitation continuum. 
Here, qualitatively different $\chitwop$ responses are found between the intermediate-field and high-field regimes, corresponding to regimes where conventional magnons break down and are restored, respectively. 
We substantiate this analysis by showing that $\chitwop$ is restricted to a distinct universal form on the level of the linear spin-wave approximation for partially-polarized magnets. 
Deviations from this form, as predicted for \rucl at intermediate field strengths, can hence be used in 2DCS experiments as direct evidence for a breakdown of the conventional magnon picture.

\vspace{0.05cm}

\itsection{Numerical Method}%
We focus on   
the zero-temperature 
second-order %
susceptibility 
\begin{align}
\chi_{\PA \PB \PC}^{2}(t,\tau+t) &= i^2 \Theta(t) \Theta(\tau) \braket{[[\PA  (\tau+t),\PB(\tau)],\PC(0)]} \nonumber%
\\
&= -2 \Theta(t)\Theta(\tau) \operatorname{Re}\Big[ \Braket{\PA e^{-i\mathcal{H}t}\PB e^{-i\mathcal H\tau} \PC}  \nonumber \\
&\quad \qquad \quad   - \Braket{\PB e^{i\mathcal{H}t}\PA e^{-i\mathcal H(t+\tau)} \PC} \Big], \label{eq:chi2_2}
\end{align}
where w.l.o.g.\ the spectrum of $\mathcal H$ was shifted  such that the ground state energy is $E_0=0$. \PA, \PB, \PC\ are operators of choice; examples are magnetization components along particular directions (relevant for terahertz 2DCS) or couplings to electrical polarization \cite{kanega2021nonlinearopticalresponsesinkitaevspinliquids,krupnitska2023finitetempSHG,brenig2024response}. %

To efficiently calculate matrix elements of the form
$
    \braket{\state{0}| \PA e^{-i\mathcal H t} \PB e^{-i\mathcal H\tau} \PC |\state{0}} \label{eq:matel}
$
appearing in \cref{eq:chi2_2}{ ($\ket{0}$ being the exact ground state)}
using Lanczos routines, we first define the startvectors %
\begin{equation}
 \ket{\phi_0^\PA}, \ket{\phi_0^\PB}, \ket{\phi_0^\PC}, \ \, \text{where}\  \ket{\phi_0^\PX} = \frac{\PX \ket{\state{0}}}{\sqrt{\braket{\state{0}|\PX^\dag \PX | \state{0}}}} = \frac{\PX \ket{\state{0}}}{N_0^\PX}.
\label{eq:NKS}
\end{equation}
For each distinct one of these, a standard Lanczos routine \cite{lanczos1950iteration} will generate a basis for the $\Krylov$-dimensional Krylov subspace
\begin{equation}
\mathrm{span}\left(
    \left\{
    \PopX\ket{0},  \mathcal H  \PopX \ket{0},  {\mathcal H}^2 \PopX \ket{0}, \dots , {\mathcal H}^{\Krylov -1}\PopX \ket{0}
    \right\}
    \right). \label{eq:Krylov}
\end{equation} 

Fixing notation, we name the $d$-dimensional ($d=\mathrm{dim}(\mathcal H)$) orthonormal basis vectors generated during the Lanczos routine as $\ket{\phi_m^\PX}$. 
Diagonalization of the  tridiagonal matrix 
yields eigenvalues $\epsilon_m^\PX$ and %
$\Krylov$-dimensional eigenvectors $v_m^\PX$%
. The latter represent the $d$-dimensional vectors  $\ket{\psi_m^\PX}=\sum_{l=0}^{\Krylov -1} v_{m,l}^\PX \ket{\phi_l^\PX}$ of the full Hilbert space.

Rewriting the first matrix element in \cref{eq:chi2_2} as
\begin{align}
&\quad \braket{\state{0}| \PA e^{-i\mathcal H t} \PB e^{-i\mathcal H\tau} \PC |\state{0}} \nonumber
    \\ &= N_{0}^\PA N_{0}^\PC \sum_{a,b=0}^\infty \frac{(-it)^a (-i\tau)^b}{a!b!}  \braket{\phi_0^\PA| {\mathcal H}^a \PB {\mathcal H}^b | \phi_0^\PC }, \label{eq:loijwqoiwej}
\end{align}
we emphasize that powers of %
{$a,b<\Krylov$} in $(\bra{\phi_0^\PA} \mathcal H^a)$ and $(\mathcal H^b\ket{\phi_0^\PC})$ can be exactly reproduced within the %
respective Krylov subspaces [\cref{eq:Krylov}]. 
Utilizing this, we insert projectors into the subspaces, $\projector^\PX = \sum_{n=0}^{\Krylov -1}\ket{\psi_n^\PX}\bra{\psi_n^\PX}$:
\begin{align}
   &  \braket{\phi_0^\PA| {\mathcal H}^a \PB {\mathcal H}^b | \phi_0^\PC }  =
    \braket{\phi_0^\PA| \ \projector^\PA {\mathcal H}^a\projector^\PA \, \PB\  \projector^\PC {\mathcal H}^b \projector^\PC \,| \phi_0^\PC} \nonumber
    \\ &\qquad \quad \qquad = \sum_{n,p=0}^{\Krylov -1} {v_{n,0}^{\PA}}\,  v_{p,0}^{\PC\,\ast} \, (\epsilon_n^\PA)^a (\epsilon_p^\PC)^b  \bra{\psi_n^\PA} \PB \ket{\psi_p^\PC}\label{eq:03jr0q93jr}
\end{align}
valid for {$a,b<\Krylov$}, 
where we used that 
matrix elements of $\mathcal H$ within a Krylov space obey $\braket{\psi_n^\PX | {\mathcal H}^a | \psi_m^\PX}=\delta_{nm}(\epsilon_n^\PX)^a$ exactly for $a<\Krylov$, even when $\ket{\psi_m^\PX}$ are not converged to eigenvectors of $\mathcal H$ yet. %

The approximation of the method is to insert \cref{eq:03jr0q93jr} into \cref{eq:loijwqoiwej} also for terms with %
$a,b\ge \Krylov$, i.e.\ $\braket{\psi_n^\PX | {\mathcal H}^a | \psi_m^\PX}\approx \delta_{nm}(\epsilon_n^\PX)^a$ for $a \ge \Krylov$. 
This insertion yields
\begin{align}
&\quad \braket{\state{0}| \PA e^{-i\mathcal H t} \PB e^{-i\mathcal H\tau} \PC |\state{0}} \nonumber
\\ &\approx N_0^\PA N_0^\PC \sum_{n,m=0}^{\Krylov -1} {v_{n,0}^{\PA}}\,  v_{m,0}^{\PC\ast} \,  e^{-i(t\epsilon_n^\PA + \tau \epsilon_m^\PC)} \braket{\psi_n^\PA | \PB | \psi_m^\PC}, \label{eq:sojfowierjw}
\end{align}
which introduced errors at the orders $t^\Krylov$ and $\tau^\Krylov$. 
Inserting \cref{eq:sojfowierjw} into \cref{eq:chi2_2} (and the analogue of \cref{eq:sojfowierjw} for the second matrix element in \cref{eq:chi2_2}), and moving to the frequency domain, we arrive at 
\begin{align}
	&\chi_{\PA \PB \PC}^{2}(\omega_t,\omega_\tau) =  \int_{-\infty}^{\infty} \int_{-\infty}^{\infty}  \mathrm dt\, \mathrm d\tau \,     \chi_{\PA \PB \PC}^{(2)}(t, \tau+t)  e^{i\omega_t^{+}\!t+i\omega_\tau^{+}\!\tau} \nonumber
	\\ &\approx \sum_{n,m=0}^{\Krylov -1} 
   \frac{\MX_{n,m}}{\big(\epsilon^{\PA\prime}_n-\omega^+ _{t}\big) \big(\epsilon^{\PC\prime}_m-\omega^+_\tau\big)}
   +\frac{\MX_{n,m}^\ast{}}{\big(\epsilon^{\PA\prime}_n+\omega^+ _{t}\big) \big(\epsilon^{\PC\prime}_m+\omega^+_\tau\big)} \nonumber
   \\
   &\qquad \qquad -\frac{\MY_{n,m}}{\big(\epsilon_m^{\PC\prime}-\epsilon^{\PB\prime}_n-\omega_t^+\big) \big(\epsilon_m^{\PC\prime} - \omega_\tau^+  \big)} \nonumber
 \\&\qquad \qquad  -\frac{\MY_{n,m}^\ast}{\big(\epsilon_m^{\PC\prime} - \epsilon^{\PB\prime}_n + \omega_t^+  \big) \big(\epsilon_m^{\PC\prime}+\omega_\tau^+ \big)}, \label{eq:chi2final_2}
\end{align}
where $\omega_t^{+}=\omega_t + i\eta$ and $\omega_\tau^{+}=\omega_\tau + i\eta$ with a broadening $\eta>0$, $\epsilon^{\PX\prime}_a=\epsilon^\PX_a - E_0$, and 
\begin{align}
    \MX_{n,m} 
    &= N_0^\PA N_0^\PC  {v_{n,0}^{\PA}}\,v_{m,0}^{\PC\ast} \sum_{l,p=0}^{\Krylov -1}v_{n,l}^{\PA\ast}\, v_{m,p}^{\PC} \braket{\phi_l^\PA | \PB | \phi_p^\PC}, \label{eq:O_nm2}
    \\ 
        \MY_{n,m} 
    &= N_0^\PB N_0^\PC  {v_{n,0}^{\PB}}\,v_{m,0}^{\PC\ast}\sum_{l,p=0}^{\Krylov -1}v_{n,l}^{\PB\ast}\, v_{m,p}^{\PC} \braket{\phi_l^\PB | \PA | \phi_p^\PC}. \label{eq:O_nm3}
\end{align} 
Note that if $\PA=\PB$, then $\MX_{n,m}=\MY_{n,m}$.

The presented method becomes exact when $\Krylov=\mathrm{dim}(\mathcal H)$, in which case the Krylov space covers the full Hilbert space and $\epsilon^\PX_n$ ($\ket{\psi^\PX_n}$) become the exact eigenenergies (eigenstates) of $\mathcal H$. In practice, we expect that in most cases $\Krylov \ll \mathrm{dim}(\mathcal H)$ will be sufficient for well-converged results of $\chi^{2}$, even without the excited Lanczos eigenstates $\ket{\psi_n^\PX}$ being converged to eigenstates of $\mathcal H$. This can be understood intuitively from the fact that the method captures all contributions to  $\chi^{2}$ up to the $\Krylov$'th order in $t,\tau$ exactly [cf.\ \cref{eq:loijwqoiwej}]. %

The method becomes computationally cheaper if %
{some of  $\PA,\PB,\PC$ are equal to another. When $\PA=\PB=\PC$, }likely the %
{most applied case}, only a single Lanczos space 
needs to be spanned. %
An algorithm implementing that case is described in Appendix~\hyperref[appendix:implementation]{A}.

\vspace{0.05cm}

\begin{figure}
    \centering
    \includegraphics[width=.8\linewidth]{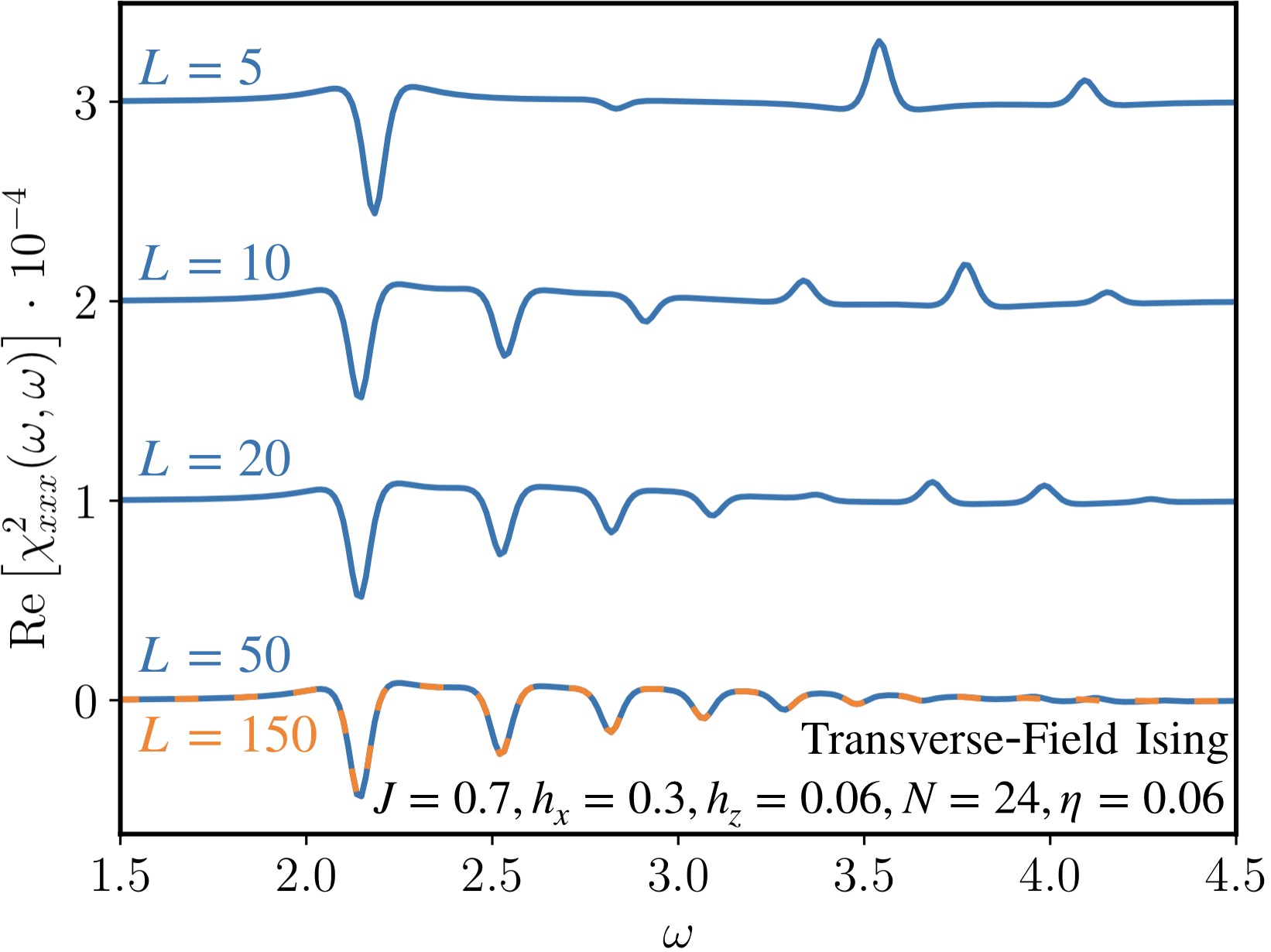}
    \caption{
    Comparison of computed $\chi^2_{xxx}(\omega_t=\omega,\omega_\tau=\omega)$ for different employed Krylov space dimensions $\Krylov$ shown for the transverse-field Ising model
    $\mathcal H=\sum_i^N -J\sigma_i^z \sigma_{i+1}^z - h_x \sigma_i^x - h_z \sigma_i^z$ with $J=0.7, h_x=0.3,h_z=0.06$ and a broadening $\eta=0.06$
    on $N=24$ sites. %
    For \cref{eq:chi2_2}, the subscript of $\chi^2_{xxx}$ denotes $\PA=\PB=\PC=\sum_i \sigma_i^x$. %
    Results for different $\Krylov$ are offset by a constant, except $\Krylov =150$, which is shown as a dashed line. 
}
    \label{fig:convergence_main}
\end{figure}

\itsection{Benchmarking}%
 We performed benchmarks against the results on the transverse-field Ising model from Ref.~\cite{watanabe2024exploring}, who simulated the two-pulse measurement protocol using explicit time evolution on a finite two-dimensional time grid with subsequent Fourier transform into frequency space.  %
At sufficiently large Krylov space dimension $\Krylov$, we find excellent agreement with   Ref.~\cite{watanabe2024exploring} throughout the two-dimensional frequency plane (see Appendix~\hyperref[appendix:benchmarks]{B}). 
Results for the frequency-plane diagonal  ($\omega_t=\omega_\tau$), which hosts the most dominant intensity features for this model, are shown in \cref{fig:convergence_main} for different values of employed $\Krylov$, demonstrating the convergence behavior. 
In \cref{fig:convergence_main}, the strongest features are already found to set in for very small $\Krylov\sim 5$ and $\Krylov\sim 10$. Satisfactory convergence sets in %
at circa $\Krylov\sim 50$, with no significant change compared to higher $\Krylov =150$ [\cref{fig:convergence_main}]. %
A similar convergence behavior was observed for the results %
discussed %
{later}. %
 For the models %
 considered %
in this study, the calculation of $\chi^2(\omega_t,\omega_\tau)$ for $\Krylov=150$ was of similar computational cost as the computation of the ground state (via \cite{lehoucq1998arpack}), making the method rather cheap.

\vspace{0.05cm}

\begin{figure}
	\centering
    \includegraphics[width=\linewidth]{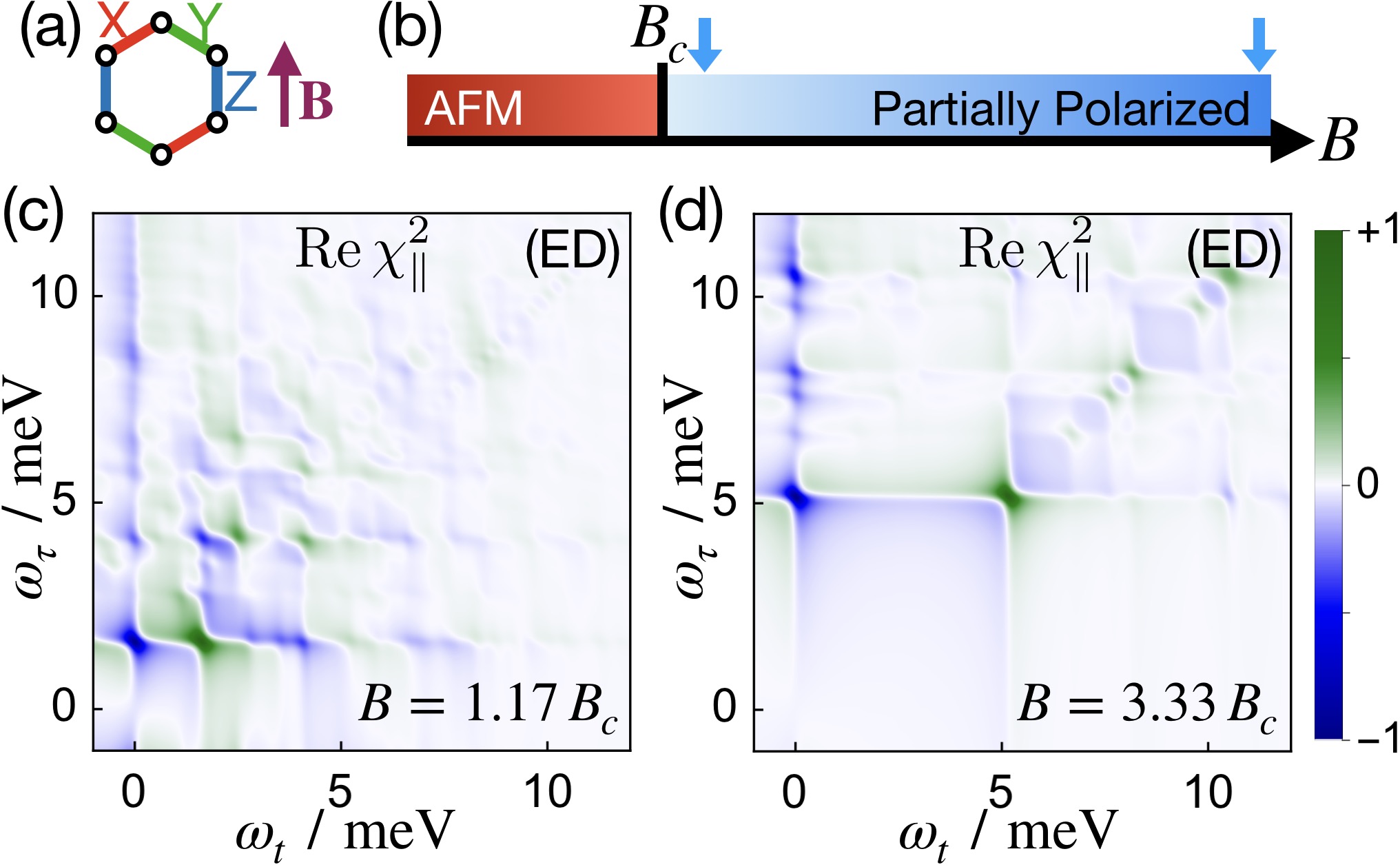}
    \caption{%
 (a) Definition of bond types %
 and chosen field direction~$\mathbf B$. 
    (b) $T=0$ Phase diagram for the considered \rucl model under in-plane magnetic field. %
    (c,d) Second-order terahertz response 
 $\mathrm{Re}\,\chitwop$  %
 {for} (c)~$B=1.17 B_c$, and (d)~$B=3.33 B_c$. Computed within ED using the presented  method with $\Krylov=150$, $\eta=0.2\,\mathrm{meV}$ on a 24-site $C_3$-symmetric periodic cluster. %
 $B_c=6\,\mathrm{T}$ within ED. The selected plot range focuses on the first frequency quadrant, which contains all main intensity features %
  except for their counterparts in the  fourth quadrant, trivially related by    $\chi^2(\omega_t,\omega_\tau) =  {\chi^2}^\ast(-\omega_t,-\omega_\tau)$. Color scales are independent for each plot.
    \label{fig:ED}
    }
\end{figure}

\noindent\itsection{Application to \rucl Model}
We now apply our numerical method to extended Kitaev models on the honeycomb lattice, described by
\begin{align}
   \mathcal H=&\sum_{{\langle i j\rangle}_\gamma} K S_{i}^{\gamma} S_{j}^{\gamma}+\Gamma\left(S_{i}^{\alpha} S_{j}^{\beta}+S_{i}^{\beta} S_{j}^{\alpha}\right) + J \mathbf{S}_{i} \cdot \mathbf{S}_{j} \nonumber
    \\ &+ \sum_{\langle\langle\langle i j\rangle\rangle\rangle} J_{3} \mathbf{S}_{i} \cdot \mathbf{S}_{j}- \sum_i \mu_\mathrm{B} \mathbf{B} \cdot \mathbf g \cdot \mathbf S_i
    ,
    \label{eq:HRucl}
\end{align}
where $\gamma={x,y,z}$ accords to the bond type X,Y,Z [\cref{fig:ED}(a)] and $\{\alpha,\beta\}=\{x,y,z\}\setminus \{\gamma\}$.   $K$ corresponds to the Kitaev coupling, $\Gamma$ to symmetric off-diagonal exchange and $J$ ($J_3$) to %
 nearest-neighbor (third-neighbor) Heisenberg coupling. $\mathbf B$ is the static %
 magnetic field and $\mathbf g$ the gyromagnetic tensor. 

We focus on the Kitaev candidate material \rucl under in-plane magnetic fields $\mathbf B \parallel (y-x)$ %
(parallel to a bond), for which we employ the minimal model from Refs.~\cite{winter17breakdown,winter2018Probing} as a representative one; $(K,\Gamma,J,J_3)=(-5,2.5,-0.5,0.5)\,\mathrm{meV}$ and $g_{\parallel}=2.3$. %
This model has been showcased previously to reproduce the unconventional \textit{linear} response of \rucl, in which  linear spin-wave theory and conventional magnons can break down \cite{winter17breakdown}. 
 While \rucl orders antiferromagnetically, an in-plane magnetic field of $B_c\approx 7\,\mathrm{T}$ suppresses this order [\cref{fig:ED}(b)]. The nature of the phase(s) and of the excitations beyond $B_c$ have been subject to significant debate due to numerous unconventional observations for $B\gtrsim B_c$, with the scenario of a field-induced Kitaev spin liquid and Majorana fermionic excitations under controversial scrutiny \cite{kasahara2018majorana,yokoi2020half,bruin2022robustness,czajka2023planar,lefranccois2023oscillations}. Nonetheless, undisputedly, for increasing field strengths $B\gg B_c$, the material asymptotically  approaches the conventional polarized state \cite{sahasrabudhe2020high}. 
 We will therefore investigate the higher-order response in the region out of antiferromagnetic order, $B>B_c$,  focussing on potential differences between the regimes $B\gtrsim B_c$ and $B\gg B_c$.

\begin{figure}
	\centering
    \includegraphics[width=\linewidth]{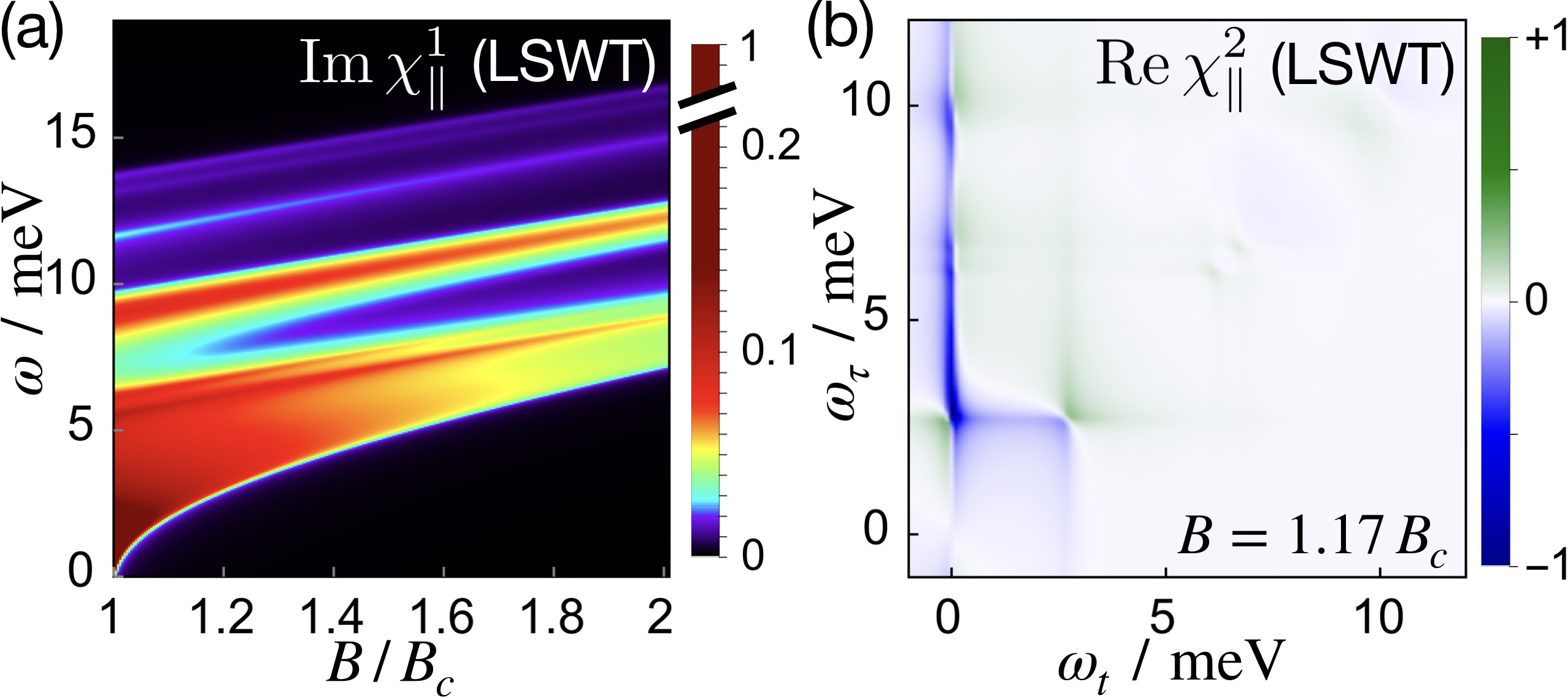}
    \caption{(a) Linear terahertz response $\mathrm{Im}\,\chionep$ at LSWT level as a function of  magnetic field for the partially-polarized phase ($B>B_c$). The intensity corresponds to the two-magnon continuum. %
    One-magnon states do not contribute in $\chionep$. %
	 (b)~Nonlinear response $\mathrm{Re}\,\chitwop$ at  LSWT level at $B=1.17B_c$.
     \label{fig:LSWT}
     }
\end{figure}

For the operators $\PA,\PB,\PC$ in $\chi^2_{\PA\PB\PC}$ [\cref{eq:chi2_2}] we choose the magnetization $\Mp$, %
corresponding to the magnetic field pulses in \cref{fig:introduction}(a) being parallel to the static external field $\mathbf B$, and to magnetic-dipole coupling with the light. 
This choice is motivated by the corresponding linear-response channel $\chionep$ featuring an excitation continuum in \rucl, that has been discussed as evidence for a QSL \cite{wang2017magnetic}, [\cref{fig:LSWT}(a), discussed later]. 
{We abbreviate $\chi^2_{\Mp\Mp\Mp}\equiv \chi^2_{\parallel}$}.  %

Exact diagonalization (ED) results of $\chitwop$ are shown in Figs.~\ref{fig:ED}(c,d) for a low-field case ($B=1.17B_c$) and a high-field case ($B=3.33B_c$), computed using the presented method with $\Krylov=150$ and $\eta=0.2\,\mathrm{meV}$ %
{on $N=24$ sites. Note that, in finite-size calculations, %
excitation continua generally appear  as series of discrete states. Similar as it is established for  finite-size simulations of \textit{linear} response \cite{dagotto1994correlated}, one could alleviate this discreteness \textit{ad hoc} by employing a sufficiently large broadening $\eta$. }{While we suspect the poles on the diagonal in \cref{fig:ED}(d) to form a continuum in the thermodynamic limit, we choose here a cautious (small) broadening $\eta=0.2\,\mathrm{meV}$ in favor of transparently presenting the new method's raw results. Whether the high-intensity pole at $\sim 5\,\mathrm{meV}$ represents the bottom of this continuum or a distinct bound state outside of the continuum \cite{sahasrabudhe2020high}, is hard to discern in finite-size calculations, but not focus of this study. } %

  We want to highlight the qualitatively different results between the regimes $B\gtrsim B_c$ and $B\gg B_c$. In the \textit{high-field} regime ($B\gg B_c$) shown in \cref{fig:ED}(d), the response is dominated by poles located on two distinct lines within the frequency plane; the frequency-diagonal ($\omega_t=\omega_\tau$, so-called ``non-rephasing signal'') and the frequency-vertical ($\omega_t=0$, $\omega_\tau\neq0$, ``rectification signal''), which we  abbreviate \FDV, respectively.  
The finite intensity away from these lines primarily stems from the broadening of their poles: Broadening arises partly from the artificial broadening $\eta$ but mostly from the natural broadening $\sim\! \frac{1}{\omega_t \omega_\tau}$ of higher-order susceptibilities, %
{related to phase twisting }\cite{khalil2003coherent,hart2023ExtractingSpinonSelfenergies,watanabe2024exploring}. %
 Contributions from distinct poles located \textit{outside} of \FDV are present but play a secondary role. 

\color{black}
In contrast, at lower fields $B\gtrsim B_c$, shown in \cref{fig:ED}(c), the majority of the intensity stems from poles that are located away from \FDV. %
Overall, these lead to an inhomogeneous continuum, spread across the two-dimensional frequency plane up to $\omega_t,\omega_\tau \lesssim 15\,$meV.  

To understand the origin of these two different responses, we analyze the type of matrix elements contributing to second-order susceptibilities in general: 
\begin{equation}
\chitwop \, \sim	\,\braket{0|\Mp|n}	\braket{n|\Mp|m}	\braket{m|\Mp|0} \label{eq:chi2matel},
\end{equation}
where $\ket{n}, \ket{m}$ are the system's excited states. %
The central matrix element $\braket{n|\Mp|m}$ carries additional information compared to linear response $\chi^1_{\parallel}\sim \left|\braket{0|\Mp|n}\right|^2$. 
Considering \cref{eq:chi2matel}, it is instructive to distinguish between contributions %
 with $n=m$ and those with $n\neq m$, each pictured in Figs.~\ref{fig:introduction}(b,c). 
In the two-dimensional frequency plane, contributions with $n=m$ [Fig.~\ref{fig:introduction}(b)] exclusively lead to poles  along  \FDV%
, as those dominant in \cref{fig:ED}(d). If poles appear away from these locations (as is the case in \cref{fig:ED}(c)), they can only stem from contributions with $n\neq m$, i.e.\ matrix elements between different excited states [Fig.~\ref{fig:introduction}(c)].  %

\vspace{0.05cm}

\itsection{$\chitwop$ {for Conventional Magnons}}%
To  put our numerical results into further context, we consider the $\chionep$ and $\chitwop$ response expected on the level of standard linear spin-wave theory (LSWT). While we will show specific results for the \rucl model, note that this discussion applies to the general LSWT response for the partially-polarized %
phase of any magnet.

Within LSWT, %
the magnetization operator $\Mp$ corresponds to a two-magnon operator, such that $\chionep$ (and $\chitwop$) probes exclusively the two-magnon continuum. 
The resulting $\chionep(\omega)$ for the discussed \rucl model is shown  as a function of $B/B_c$ in \cref{fig:LSWT}(a) \footnote{Note that the corresponding LSWT plot in Fig.~3(k) of Ref.~\cite{winter2018Probing} shows no intensity for $B>B_c$, as only one-magnon states were considered there.}. With increasing field strength, the two-magnon gap grows monotonically. %

Turning to $\chitwop$, 
from inspecting the form of $\Mp$ in terms of magnon operators [Appendix~\hyperref[appendix:LSWT]{C}], we find %
that contributions with $n\neq m$ [Fig.~\ref{fig:introduction}(c)] are \textit{generally} strongly suppressed on the LSWT level: 
 The only finite $n\neq m$ contributions relate to inter-magnon-band processes, which we conjecture to generally have %
tiny intensity compared to $n=m$ contributions, based on the fact that for Bravais lattices such contributions are completely forbidden, and one does not expect a qualitative difference between Bravais and non-Bravais. One non-Bravais example confirming this %
is discussed in the following.

LSWT results for the \rucl model on the honeycomb lattice %
are shown in \cref{fig:LSWT}(d) for $B=1.17B_c$.
 As described before, the dominance of the matrix elements from Fig.~\ref{fig:introduction}(b) leads to poles appearing only  on \FDV. %
  This overall form therefore represents the general form expected for the two-magnon continuum in $\chitwop$ in the partially-polarized phase of any magnet. For the present model at $B/B_c=1.17$, the bottom of the two-magnon continuum is at $\sim$$2.6\,\mathrm{meV}$ [cf.~\cref{fig:LSWT}(c)], leading in \cref{fig:LSWT}(d) to the onset of the strong rectification signal at $(\omega_t,\omega_\tau)=(0,2.6)\,\mathrm{meV}$ and a characteristic node feature at $(\omega_t,\omega_\tau)=(2.6,2.6)\,\mathrm{meV}$. LSWT results at other field strengths retain this form but with accordingly shifted energies [cf.~\cref{fig:LSWT}(a)]. %
 
 Deviations from a shape of predominantly \FDV %
 poles 
 in a measured $\chitwop$ can hence be used to directly diagnose continua to arise from unconventional excitations, where %
 {standard LSWT} does not capture the full physics. Such a case is found in our numerical ED results for \rucl at $B\gtrsim B_c$ in \cref{fig:ED}(c), where the dominant intensity arises from non-\FD and non-\FV %
 poles. As the high-field limit $B\gg B_c$ suppresses quantum fluctuations and restores conventional magnons, the according high-field result of the same model [\cref{fig:ED}(d)] recovers the LSWT-expected form of dominant \FDV poles. %
 $\chitwop$ therefore offers a direct measurement of the breakdown of conventional magnon excitations %
 away from the high-field limit in \rucl.
 Whether this unconventional response {is caused by}, e.g., decaying and interacting magnons or fractionalized excitations %
 is an %
 open question for this class of %
 materials 
 and goes beyond the scope of the current study.

\begin{table}
\renewcommand{\arraystretch}{1.2}
  \resizebox{\columnwidth}{!}{
  \begin{tabular}{|c|c|c|}
  \hline
State \& Fluctuations & Linear   $\chionep(\omega)$ & Nonlinear $ \chitwop(\omega_t,\omega_\tau)$  \\
   \hline \hline
\makecell{fully polarized\\ (no quantum fluctuations)}    & zero & zero \\
\hline
\makecell{partially polarized,\\ 
LSWT-type fluctuations
}   
& continuum & \makecell{homogeneous continuum\\ along 
\FDV
}  \\
\hline
\makecell{partially polarized, \\ %
non-LSWT
 fluctuations}
 & \makecell{continuum*} & \makecell{inhomogeneous\\ continuum*}
\\ \hline
  \end{tabular}
  }
  \caption{%
  General expected response for different states in linear %
  and nonlinear %
terahertz   response  in the high-field phase of a frustrated magnet, for the channel with the light's magnetic field parallel to the ordered moment ($\parallel$). %
*Note that the types of possible 
unconventional 
fluctuations are diverse, and can (dependent on the {material}) lead to additional features, such as %
bound states appearing outside of the continuum. %
  }
  \label{table_general} 
\end{table}

Beyond \rucl, a similar analysis can be applied to 2DCS measurement results on the high-field phases of other frustrated magnets. A{ useful summary for the interpretation of such measurements} is presented in  
 \cref{table_general}. All three cases are represented in the discussed \rucl model under in-plane fields, where the table's rows correspond to $B\rightarrow \infty$, 
$B\gg B_c$ and $B\gtrsim B_c$, respectively. 

\vspace{0.05cm}

\itsection{Outlook}%
We showed that nonlinear spectroscopy can unveil %
 crucial insights about the nature of excitations in highly frustrated spin systems: 
 By analyzing the positions of poles in the two-dimensional frequency plane of the susceptibility $\chitwop$ it is possible to directly asses the breakdown of conventional magnon excitations. 
 We predicted such an unusual $\chitwop$ response to be observable in \rucl in the highly discussed region of in-plane magnetic fields $B\gtrsim 7\,\mathrm{T}$. %
 Experimental 2DCS measurements on \rucl and on other frustrated magnets are highly desirable. %
 
With the newly introduced efficient numerical method the calculation of such response functions is straightforward and can be applied to different classes of models and materials, %
 as well as to operators beyond magnetization, for example to study nonlinear susceptibilities via coupling to the electric field of the light. %

\vspace{.5cm}

 \itsection{Acknowledgments}%
Special thanks goes to Axel Fünfhaus, Andreas Rückriegel, Peter Kopietz,  and P.~Peter Stavropoulos for helpful comments and discussions. 
We also thank Peter Armitage, Wolfram Brenig, Manfred Fiebig, Ciaran Hickey, Johannes Knolle, Alex Wietek and Stephen M. Winter, %
for fruitful  discussions. %
We gratefully acknowledge support by the {Deutsche Forschungsgemeinschaft (DFG, German Research Foundation) for funding through TRR 288—422213477 (project A05) and CRC 1487—443703006 (project A01).}

\bibliography{Notes}
 \clearpage
\widetext
\appendix
\begin{center}
\textbf{\large Appendix} \bigskip
\end{center}

\twocolumngrid

\setcounter{equation}{0}
\setcounter{figure}{0}
\setcounter{table}{0}
\makeatletter
\renewcommand{\theequation}{A\arabic{equation}}
\renewcommand{\thefigure}{A\arabic{figure}}
\renewcommand{\thetable}{A\Roman{table}}

\section{Appendix A: Numerical implementation\label{appendix:implementation}}
We explain the algorithm for the case of diagonal susceptibilities $\chi^{2}_{\PA\PA\PA}$, i.e.\ $\PA=\PB=\PC$ in \cref{eq:chi2_2}. %
Then 
the method to compute $\chi^{2}_{\PA\PA\PA}$ can be implemented follows:
\begin{enumerate}
    \item %
Compute the ground state $\ket{0}$ of $\mathcal H$ and its energy $E_0$, for example via a standard Lanczos routine \cite{lanczos1950iteration} using a random start vector, or related methods~\cite{lehoucq1998arpack}. 
    \item 
Generate $\ket{\phi_0^\PA}$ and $N_0^\PA$ according to \cref{eq:NKS}. 
    \item 
Using the Lanczos algorithm with $\ket{\phi_0^\PA}$ as a start vector, generate and store 
    the basis vectors $\{\ket{\phi_0^\PA},\ket{\phi_1^\PA},\dots,\ket{\phi_{\Krylov -1}^\PA}\}$ as well as the eigenvalues $\epsilon_m^\PA$ and eigenvectors $v_m^\PA$ of the tridiagonal matrix. 
    \item 
Compute all matrix elements $\braket{\phi_l^\PA | \PA | \phi_p^\PA}$ for $l,p\in\{0,1,\dots,\Krylov -1\}$. For this, it might be efficient to iterate over $p$, generating $\ket{{\phi_p^\PA}'}= \PA\ket{\phi_p^\PA}$ and computing the overlaps $\braket{\phi_l^\PA | {\phi_p^\PA}'}=\braket{\phi_l^\PA | \PA | \phi_p^\PA}$ %
    for all $l\le p$. 
    For $\PA^\dag=\PA$,  the elements with $l>p$ follow via $\braket{\phi_l^\PA | \PA | \phi_p^\PA} = (\braket{\phi_p^\PA | \PA | \phi_l^\PA})^\ast$.
    \item 
Obtain all $\MX_{n,m}$ via \cref{eq:O_nm2}. Note that the sum in \cref{eq:O_nm2} can be efficiently computed by expressing it as a matrix multiplication.%
    \item 
Evaluate $\chi_{\PA\PA\PA}^{(2)}(\omega_t,\omega_\tau)$ (here, $\MY_{n,m}=\MX_{n,m}$) via  \cref{eq:chi2final_2} with a chosen broadening $\eta>0$ %
    for all desired frequencies $\omega_t,\omega_\tau$. %
\end{enumerate}

The $d$-dimensional eigenvectors in the Krylov subspace ($\ket{\psi_m}$) do {not} need to be assembled explicitly at any point. 
The computationally expensive steps in the method are (aside from step~1, which depends on the method of choice), the steps~3 and 4, where in step~3 the Hamiltonian has to be applied $\Krylov$-times, and in step~4 one has to apply $\PB$ $\Krylov$-times and calculate %
$\frac{\Krylov^2+\Krylov}{2}$
overlaps. Depending on the choice of $\Krylov$, step~4 can therefore become the most costly step and effectively limit the range of feasible $\Krylov$.  
We note that within our numerical simulations so far, $\chi^{2}$ appears to converge for a given model at $\Krylov$ of similar sizes as similar Lanczos-based methods such as those for linear response~\cite{dagotto1994correlated} or the finite-temperature Lanczos method~\cite{PhysRevB.49.5065,jaklivc2000finite}, which are often used with %
{$50\lesssim \Krylov \lesssim 150$}. %

\section{Appendix B: Benchmarks\label{appendix:benchmarks}}
\begin{figure}
    \centering
    \includegraphics[width=1\linewidth]{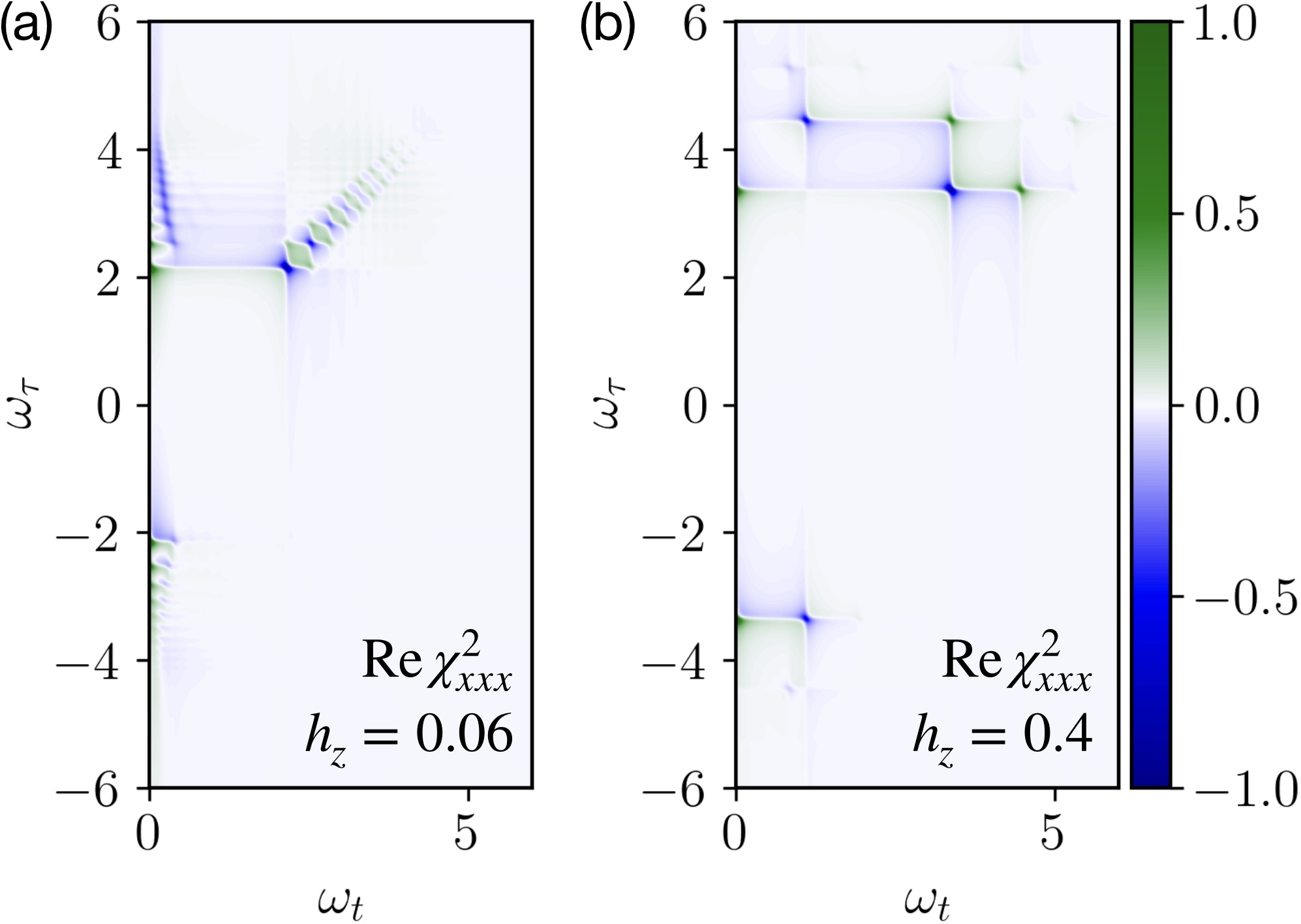}
    \caption{$\mathrm{Re}\,\chi^{2}_{xxx}(\omega_t, \omega_{\tau})$ computed using the presented algorithm with $L=150, \eta=0.06$. (a) $J = 0.7$, $h_x = 0.3$, $h_z = 0.06$ in \cref{eq:app:Ising}, (b) $J = 0.7$, $h_x = 0.3$, $h_z = 0.4$. (a,b)~can be compared to Fig.~6(b) and Fig.~6(d) in Ref.~\cite{watanabe2024exploring}, respectively.}
    \label{fig:CiaranBoth}
\end{figure}

We benchmarked our method with the transverse-field Ising model (TFIM) against the numerical study of Ref.~\cite{watanabe2024exploring}, that is based on explicit time evolution and subsequent Fourier transform in ED. The Hamiltonian of the TFIM is given by
 \begin{align}
    \mathcal H=\sum_i^N \left(-J\sigma_i^z \sigma_{i+1}^z - h_x \sigma_i^x - h_z \sigma_i^z \right)\label{eq:app:Ising}
\end{align}
with the Pauli matrices $\sigma_i^\gamma$ and the coupling constant $J$ as well as the field in transverse (longitudinal) $x$ ($z$) direction $h_x$ ($h_z$) with $J,h_x>0$. We employ the same cluster size ($N=24$), and compute in our method the susceptibility $\chi^2_{xxx}(\omega_t,\omega_\tau)$, corresponding to $\PA=\PB=\PC=\sum_i^N \sigma_i^x$ in \cref{eq:chi2_2}. 
Results for $\Krylov=150$ are shown for two parameter sets (described in the figure captions) in \cref{fig:CiaranBoth}(a) and \cref{fig:CiaranBoth}(b), which can be compared to panels within Fig.~6(b) and Fig.~6(d) of Ref.~\cite{watanabe2024exploring}, respectively. We find excellent agreement with their results.

\newcommand{\NumberSublattices}{\ensuremath{{Z}}}
\newcommand{\myC}{\ensuremath{{C}}}
\newcommand{\Cone}{\ensuremath{{U}}}
\renewcommand{\myC}{\Cone}
\newcommand{\Ctwo}{\ensuremath{{V}}}
\newcommand{\magEnergy}{\ensuremath{{\epsilon}}}

\section{Appendix C: Linear spin-wave theory details\label{appendix:LSWT}}
We consider standard linear spin-wave theory (LSWT) using the  Holstein-Primakoff expansion \cite{holstein1940field} and assume a field-polarized  ground state (all moments parallel to magnetic field $\mathbf B$). 
In this framework, for a lattice with $\NumberSublattices$ sites per unit cell, the magnetization component %
parallel to the magnetic field, $\Mp$ ($=S^z_{\mathbf q=0}$ in the standard laboratory frame), %
is given by
\begin{align}
   \Mp   &= NS - 
   \sum_{k}\sum_{s}^\NumberSublattices a^\dag_{\mathbf{k},s}a_{\mathbf{k},s} ,\label{eq:MHolstein}
   \end{align}
   where $a_{\mathbf{k},s}$ are the Holstein-Primakoff bosons of sublattice $s$ at momentum $\mathbf k$, $N$ the number of sites and $S$ the spin length. 

   Consider a generalized Bogoliubov transformation %
   that diagonalizes the LSWT Hamiltonian of question, 
   \begin{align}
       a_{\mathbf{k}s}=\sum_{l}^{\NumberSublattices} \left(\myC_{\mathbf{k}sl} \,m_{\mathbf{k}l} + \Ctwo_{\mathbf{k}sl}\, m^\dag_{-\mathbf{k}l}\right),\label{eq:Bogol}
   \end{align}
   where $m_{\mathbf{k}l}$ is the Bogoliubov quasiparticle of the $l$'th magnon band at momentum $\mathbf k$.

With \cref{eq:MHolstein,eq:Bogol}, the magnetization becomes
   \begin{align}
   \Mp &= NS -  \sum_{\mathbf{k}} \sum_{slb}^\NumberSublattices 
   \left(\myC^\ast_{\mathbf{k}sl}\, m^\dag_{\mathbf{k}l} + \Ctwo^\ast_{\mathbf{k}sl}\, m_{-\mathbf{k}l} \right) \nonumber
   \\ &\qquad \qquad \qquad \quad  \times 
   \left( \myC_{\mathbf{k}sb}\, m_{\mathbf{k}b} + \Ctwo_{\mathbf{k}sb}\, m_{-\mathbf{k}b}^\dag  \right)
   . \label{eq:M_Bogol_nonBravais}
\end{align}

Turning to dynamical response functions, the linear-order response in the channel we focus on, \chionep, is defined as 
\begin{equation}
    \mathrm{Im}\,\chionep(\omega) = \sum_n \left|\braket{n|\Mp | 0}\right|^2 \,\left[\delta(E_n-\omega)-\delta(E_n+\omega)\right],
    \label{eq:chionep_def_appendix}
\end{equation}
where $E_n$ ($\ket{n}$) is the $n$'th eigenenergy (eigenstate) of the LSWT Hamiltonian and $\delta(x)$ the Dirac delta function. With \cref{eq:M_Bogol_nonBravais}, it follows that the accessed excited states  in $\chionep$ are two-magnon states $\ket{\mathbf{k}l,-\mathbf{k}b}=m^\dag_{\mathbf{k}l}m^\dag_{-\mathbf{k}b}\ket{0}$ with energy $E_n=\magEnergy_{\mathbf{k}l}+\magEnergy_{-\mathbf{k}b}$ and corresponding matrix element $\braket{\mathbf{k}l,-\mathbf{k}b|\Mp|0} = -\sum_s^{\NumberSublattices}\myC^\ast_{\mathbf{k}sl}\Ctwo_{\mathbf{k}sb}$. Hence, the two-magnon continuum is probed in $\chionep$, as shown for the discussed extended Kitaev model in \cref{fig:LSWT}(a). 

Considering \cref{eq:chi2matel}, the $\chitwop$ response then additionally probes matrix elements $\braket{n|\Mp|m}$ between two-magnon states $\ket{n}=\ket{\mathbf{k}l,-\mathbf{k}b}, \ket{m}=\ket{\mathbf{k}'l',-\mathbf{k}b'}$. 
With \cref{eq:M_Bogol_nonBravais}, this yields for $n=m$ contributions [\cref{fig:introduction}(b)]:
    \begin{align}
     &       \braket{\mathbf{k}l,-\mathbf{k}b|\Mp|\mathbf kl,-\mathbf{k}b}    \nonumber
     = %
     { NS - \sum_\mathbf{k}\sum_{sl}^Z |V_{\mathbf{k}sl}|^2}%
    \\&\qquad - \sum_{s}^\NumberSublattices \left( |\myC_{\mathbf{k}sl}|^2+|\myC_{-\mathbf{k}sb}|^2 + |\Ctwo_{\mathbf{k}sb}|^2 + |\Ctwo_{-\mathbf{k}sl}|^2 \right).
    \end{align}

For $n \neq m$ contributions $\braket{\mathbf{k}'l',-\mathbf{k}'b'|\Mp|kl,-\mathbf{k}b}$ [\cref{fig:introduction}(c)], it follows from the form of \cref{eq:M_Bogol_nonBravais} that the matrix element is nonzero only  if $k=k'$ \footnote{We restrict $\mathbf k$ and $\mathbf k'$ to one half of the Brillouin zone, in order to not double-count identical two-magnon states $\ket{\mathbf{k}l,-\mathbf kl}=\ket{-\mathbf{k}l,\mathbf kl}$.} %
and either $l=l'\wedge b\neq b'$  or $l\neq l' \wedge b=b'$; i.e.\ matrix elements where exactly one magnon switches into a different band.  
It follows directly, that for Bravais lattices ($\NumberSublattices=1$) there are no $n\neq m$ contributions, as there is only one magnon band. 
For non-Bravais lattices, the  $n\neq m$ contributions to $\chitwop$ are given by
\begin{align}
    \braket{\mathbf{k}l,-\mathbf{k}b'|\Mp|\mathbf kl,-\mathbf{k}b}   &= -\sum_s^\NumberSublattices \left(\myC^\ast_{-\mathbf{k}sb'}\myC_{-\mathbf{k}sb} + \nonumber \Ctwo^\ast_{\mathbf{k}sb}\Ctwo_{\mathbf{k}sb'}\right) , \\
    \braket{\mathbf{k}l',-\mathbf{k}b|\Mp|\mathbf kl,-\mathbf{k}b}   &= -\sum_s^\NumberSublattices \left(\myC^\ast_{\mathbf{k}sl'}\myC_{\mathbf{k}sl} + \Ctwo^\ast_{-\mathbf{k}sl}\Ctwo_{-\mathbf{k}sl'}\right),
\end{align}
where $b'\neq b$ and $l'\neq l$, respectively.

\bigskip

The discussion up to this point is valid for any LSWT Hamiltonian with field-polarized ground state. To obtain the explicit results on the honeycomb-lattice extended Kitaev model shown in \cref{fig:LSWT}, we performed standard LSWT  for the model of \cref{eq:HRucl},  obtaining the magnon eigenenergies and the coefficients $U_{\mathbf{k}sl}, V_{\mathbf{k}sl}$. Detailed descriptions of the LSWT for extended Kitaev models can be found, for example, in Refs.~\cite{vladimirov2016magnetic,maksimov2020rethink,smit2020magnon}. 

$\chionep(\omega)$ [\cref{fig:LSWT}(a)] was obtained by evaluating \cref{eq:chionep_def_appendix} on a $k$-grid of 40\,000 points and a Lorentzian broadening of $0.1\,$meV for a range of $B_c<B<2B_c$, where $B_c\approx11\,\mathrm{T}$ within LSWT for the chosen model and field direction. 

$\chitwop(\omega_t,\omega_\tau)$ [\cref{fig:LSWT}(b)] was obtained by evaluating 
\begin{equation}
 \chitwop%
  = \sum_{n,m}	\braket{0|\Mp|n}	\!\braket{n|\Mp|m}\!	\braket{m|\Mp|0}  g(E_n, E_m,\omega_t,\omega_\tau),\label{eq:app:totalchi2p}
\end{equation}
where the sum $\sum_{n,m}$ goes both over the ground state and the excited states, and 
\begin{align}
g(E_m,E_n,\omega_t,\omega_\tau) 
&=\tfrac{2 E_m-E_n}{(E_m+{\omega_t^+}) (E_n+{\omega_\tau^+}) (E_m-E_n-{\omega_t^+})} \nonumber
\\&\quad +\tfrac{E_m-2 E_n}{(E_m-{\omega_\tau^+}) (E_n-{\omega_t^+}) (E_m-E_n-{\omega_t^+})}
\end{align}
using the preceding expressions for the matrix elements and the same $k$-grid and broadening. 

In these calculations, the largest summed contributions of the type in \cref{fig:introduction}(c) (i.e.\ $n\neq m$ contributions where $n,m$ are  excited states) throughout the two-dimensional frequency-plane were at least two orders of magnitudes smaller than the contributions of the type in \cref{fig:introduction}(b) ($n=m$). This leads to the form of essentially dominant \FDV poles in $\chitwop(\omega_t,\omega_\tau)$, as discussed in the main text.

\end{document}